\documentclass{iau} 
\usepackage{graphicx}
\usepackage{natbib}
\usepackage{wrapfig}

\usepackage{changes}    
\usepackage{comment}    
\usepackage{xcolor}     

\def\beq#1{\begin{equation}\label{#1}}

\def\eeq{\end{equation}}

\def\beqa#1{\begin{eqnarray}\label{#1}}

\def\eeqa{\end{eqnarray}}

\def\Eq#1{Eq.~(\ref{#1})}

\def\eqn#1{~(\ref{#1})}

\def\myfrac#1#2{\left(\frac{#1}{#2}\right)}

\def\mycomment#1{\relax}

\title[Accreting neutron stars in HMXBs] 
{X-ray binaries with neutron stars at different accretion stages}

\author[K. Postnov, A. Kuranov \& L. Yungelson]   
{Konstantin A. Postnov$^{1,2}$, Alexander G. Kuranov$^{1,3}$ \& Lev R. Yungelson$^4$}

\affiliation{$^1$Sternberg Astronomical Institute, Moscow M.V. Lomonosov State University\\ 
13 Universitetskij pr., 119234 Moscow, Russia\\
email: {\tt pk@sai.msu.ru} \\[\affilskip]
$^2$Kazan Federal University, 
18 Kremlyovskaya str., 420008 Kazan, Russia
\\[\affilskip]
$^3$Russian Foreign Trade Academy,  4а Pudovkin str., 119285 Moscow, Russia
\\[\affilskip]
$^4$Institute of Astronomy, RAS,
48 Pyatnitskaya str., 119017 Moscow, Russia
\\email: {\tt lry@inasan.ru}}

\pubyear{2019}
\volume{346}  
\jname{High-mass X-ray binaries:
illuminating the passage from massive binaries  to merging compact objects}
\editors{L. Oskinova, E. Bozzo \& T. Bulik, eds.}
\begin{document}

\maketitle

\begin{abstract}
Different accretion regimes onto magnetized NSs in HMXBs are considered: wind-fed supersonic (Bondi) regime at high accretion rates $\dot M\gtrsim 4\times 10^{16}$~[g s$^{-1}$] , subsonic settling regime at lower $\dot M$ and supercritical disc accretion during Roche lobe overflow. In wind-fed stage, NSs in HMXBs reach equilibrium spin periods $P^*$ proportional to binary orbital period $P_b$.  At supercritical accretion stage, the system may appear as  a pulsating ULX.  Population synthesis of Galactic HMXBs using standard assumptions on the binary evolution and NS formation is presented. Comparison of the model $P^*$ -- $P_b$ (the Corbet diagram), $P^*$ -- $L_x$ and $P_b$ -- $L_x$ distributions with those for the observed HMXBs (including Be X-ray binaries) and pulsating ULXs suggests the importance of the reduction of $P^*$ in non-circular orbits, explaining the location of Be X-ray binaries in the model Corbet diagram, and the universal parameters of pulsating ULXs depending only on the NS magnetic fields.

\keywords{accretion, accretion disks; binaries: close; stars: neutron}
\end{abstract}

\firstsection 
\section{Introduction}
High-mass X-ray binaries (HMXBs) are important tools to study binary evolution (see other papers in this volume). Here we focus on two particular aspects of accretion onto magnetized neutron stars (NSs) in HMXBs that have important observational manifestations: wind accretion in the settling (subsonic regime), which is relevant at moderate and low accretion rates onto NS, $\dot M\lesssim 4\times 10^{16}$~[g s$^{-1}$]. For wind accretion in HMXBs with $\dot M >4\times 10^{16}$ [g s$^{-1}$], the supersonic Bondi-Hoyle-Littleton accretion takes place.
We separately consider HMXBs with supercritical disc accretion at $\dot M\gtrsim 10^{18}$~[g s$^{-1}$], which is relevant to the growing class of pulsating ultraluminous X-ray sources (ULXs) (see F. Harrison and 
F. F\"{u}rst papers in this volume). 
\label{s:intro}

\section{Settling wind accretion in HMXBs: an overview}
\label{s:theory}

The settling accretion theory is 
designed to describe the interaction of accreting plasma with magnetospheres of slowly 
 rotating NSs (see \citet{APA} for a detailed derivation and discussion). 
The key feature of the model is the calculation of the plasma entry rate into the magnetosphere due to the Rayleigh-Taylor 
 (RT) instability regulated by plasma cooling (Compton or radiative). 
RT instability can be suppressed by the fast rotation of the magnetosphere \citep{1980ApJ...235.1016A}, and settling
accretion regime can be realized for NS spin periods
$
P^*>27 [\mathrm{s}] \dot M_{16}^{1/5}\mu_{30}^{33/35}(M_x/M_\odot)^{-97/70}.
$
At shorter
 NS periods, accretion regime at any X-ray luminosity will be supersonic because of efficient plasma penetration into the magnetosphere via 
Kelvin-Helmholtz instability \citep{1983ApJ...266..175B}.
Here and below, accretion rate  $\dot{M}=\dot{M}_{16}\times 10^{16}[\mathrm{g\,s}^{-1}]$, NS magnetic moment $\mu=\mu_{30}\times 10^{30}[\mathrm{G\,cm}^3]$.  Compton cooling of accreting plasma by X-ray photons generated near the NS surface enables a free-fall (Bondi-type) supersonic accretion onto the NS magnetosphere provided that the X-ray luminosity is above the critical value $L_{x}^\dag\simeq 4\times 10^{36}$ erg s$^{-1}$ \citep{2012MNRAS.420..216S}, corresponding to the mass accretion rate onto  NS $\dot M_x\simeq 4\times 10^{16}$~[g s$^{-1}$]. At lower X-ray luminosities, the plasma above the magnetosphere remains hot enough to 
avoid an effective inflow into the magnetosphere 
due to RT instability, and the plasma entry rate is controlled by the cooling rate. This results in the formation of a hot, convective shell above the magnetosphere.
In this shell, the plasma gravitationally captured by the neutron star from the stellar wind of the companion (basically, at the Bondi-Hoyle-Littleton rate, $\dot M_B$)  settles towards the NS magnetosphere at a rate $\dot M_x=f(u)\dot M_B$, with the dimensionless factor $f(u)\lesssim 0.5$\footnote{In a radial accretion flow with an account of the Compton cooling, the value $f(u) \gtrsim 0.5$ also corresponds to the location of the sonic point  below the magnetospheric boundary $R_A$ \citep{2012MNRAS.420..216S} enabling a free-fall plasma flow down to 
$R_A$ and the formation of a shock above the magnetosphere 
\citep{1976ApJ...207..914A}.}, whose precise value depends on the plasma cooling regime (Compton or radiative). With a good accuracy, this factor can be written as 
$
f(u)\approx (t_\mathrm{ff}/t_\mathrm{cool})^{1/3}\!,
$
where $t_\mathrm{ff}$ is the free-fall time at the magnetospheric boundary (the Alfv\'en radius, $R_A$), $t_\mathrm{cool}$ is the characteristic plasma cooling time. Clearly, in the case $t_\mathrm{cool}\gg t_\mathrm{ff}$, this factor can be very small, leading to an effective decrease in the mass accretion rate onto the NS surface compared to the maximum possible Bondi rate, $\dot M_B$. 

In a circular binary system with the components separation $a$,  Bondi-Hoyle-Littleton accretion rate can be estimated as
\beq{e:dotMB}
\dot M_B\approx \frac{1}{4}\dot M_\mathrm{o}\frac{v}{v_w}\myfrac{R_B}{a}^{2}\!,
\eeq
where $\dot M_\mathrm{o}$ is the optical star wind mass-loss rate,  gravitational Bondi radius 
$
R_B=\delta\frac{2GM_x}{v^2},
$
$M_x$ -- NS mass,  $v^2=v_\mathrm{orb,x}^2+v_w^2$ -- relative stellar wind velocity, $v_\mathrm{orb,x}$ -- NS orbital velocity. The estimate 
\eqn{e:dotMB} assumes a spherically-symmetric wind from the optical star and is applicable for $R_B\ll a$. Numerical factor $\delta\sim 1$ 
takes into account the actual location of the bow shock in the stellar wind \citep{1971MNRAS.154..141H}. Modern numerical calculations \citep[see, e.g.,][]{2017ApJ...846..117L} suggest smaller (up to an order of magnitude) mass accretion rates onto a gravitating mass from the stellar wind than given by the standard Bondi-Hoyle-Littleton formula, which can be reformulated in terms of smaller values of the parameter $\delta\simeq 0.3-0.5$.
Other studies \citep[e.g.,][]{2017MNRAS.468.3408D} claim that accretion rate can exceed the Bondi one due to focusing of the wind. Therefore, the numerical factor $\delta$ 
can differ from unity by a factor of a few.

The Alfv\'en radius is defined from the pressure balance between the accreting plasma and the magnetic field at the magnetospheric boundary and depends on the actual mass accretion rate, $\dot M_x$ (i.e., the observable X-ray luminosity $L_x$), and the neutron star magnetic moment, $\mu$:
$R_A\sim (f(u)\mu^2/(\dot{M}_x\sqrt{GM_x}))^{2/7}\!\!.
$
Factor $f(u)$ is  
\beq{e:fuC}
f(u)_\mathrm{Comp}\approx 0.22 \zeta^{7/11} \dot M_{x,16}^{4/11}\mu_{30}^{-1/11}
\eeq
and 
\beq{e:furad}
f(u)_\mathrm{rad}\approx 0.1 \zeta^{14/81} \dot M_{x,16}^{6/27}\mu_{30}^{2/27}
\eeq
for the Compton and radiative cooling, respectively \citep{2017arXiv170203393S,APA}. Here $\zeta\lesssim 1$ is the numerical factor determining the characteristic scale of the growing RT mode (in the units of the Alfv\'en radius $R_A$).  

Thus, specifying the plasma cooling mechanism, we are able to estimate the expected reduction in the mass accretion rate at the settling accretion stage, $f(u)=F(\dot M_x,...)=F(f(u)\dot M_B,...)$, and by solving for $f(u)$, to find the explicit expression for $\dot M_x$ as a function of $\dot M_B$ and other parameters.

Putting all things together, we are able to express the expected accretion rate onto the NS (which can be directly probed by the observed X-ray luminosity $L_x$) 
using the Bondi capture rate $\dot M_B$, which can be calculated 
from the known mass-loss rate of the optical companion $\dot M_o$, stellar wind velocity $v_w$ and binary system parameters:
\beq{e:MxC}
\dot M_{x,16}^\mathrm{Comp}\simeq 0.1 \zeta \dot M_{B,16}^{11/7}\mu_{30}^{-1/7}
\eeq
for the Compton cooling and 
\beq{e:Mxrad}
\dot M_{x,16}^\mathrm{rad}\simeq 0.05\zeta^{2/9}
\dot M_{B,16}^{9/7}\mu_{30}^{2/21}
\eeq
for the radiative cooling. The actual accretion rate onto NS is taken to be $\dot M_x=\max\{\dot M_{x}^\mathrm{Comp},\dot M_{x}^\mathrm{rad}$\} and determines the cooling regime. Matching \Eq{e:MxC} and \Eq{e:Mxrad} shows that for a given NS magnetic momentum, the Compton cooling of plasma dominates for the gravitational capture mass rate
$\dot M_{B,16}\gtrsim 0.1 \zeta^{-2/9}\mu_{30}^{5/6}\,.
$


During the settling accretion stage, a hot convective shell formed above the NS magnetosphere mediates the angular momentum transfer to/from the rotating NS enabling long-term spin-down episodes with spin-down torques correlated with the X-ray luminosity, as observed, for example, in GX 1+4 \citep{1997ApJ...481L.101C,2012A&A...537A..66G}. 
Turbulent stresses acting in the shell lead to an almost iso-momentum angular velocity radial distribution, $\omega(r)\sim 1/r^2$, suggesting the conservation of 
the specific angular momentum of gas captured near the Bondi radius $R_B$, $j_w = \eta \omega_b R_B^2$, $\eta\approx 1/4$ \citep{1975A&A....39..185I}, where $\omega_b=2\pi/P_b$ is the orbital angular frequency. 
The numerical coefficient $\eta$ here can vary in a wide range due to inhomogeneities in the stellar wind and can be even negative \citep{1989MNRAS.238.1447H} leading to a temporal formation of retrograde accretion discs . In our calculations, we varied this parameter in the range $\eta=[0,0.25]$.

The condition for a quasi-spherical accretion to occur is that the specific angular momentum of a gas parcel is smaller than the Keplerian angular momentum at the Alfv\'en radius: $j_w\le j_K(R_A)=\sqrt{GM_xR_A}$. In the opposite case, $j_w>j_K(R_A)$, the formation of an accretion disc around the magnetosphere is possible. 

The equation for spin evolution of a NS at the settling accretion stage can be written in the form \citep{2012MNRAS.420..216S,APA}
\beq{e:spinev}
\frac{dI\omega^*}{dt}=\eta Z\dot M_x \omega_b R_B^2-Z(1-z/Z)\dot M_x \omega^*R_A^2,
\eeq
where $I$ is the NS momentum of inertia, the coupling coefficient of the plasma-magnetosphere interaction is $Z=1/f(u)(u_c/u_{ff})$, $u_c/u_{ff}\lesssim 1$ is the ratio of the convective velocity in the shell and the free-fall velocity, $z\lesssim 1$  is  a  numerical  coefficient  which  takes  into  account  the
angular  momentum  of  the  falling  matter at the NS surface.
 The account of the variable specific angular momentum of the gravitationally captured stellar wind leads to the correction of the equilibrium NS period derived in \cite{2012MNRAS.420..216S,APA} by the factor $0.25/\eta$. In the possible case of a negative value of $\eta$,
NS would rapidly spin-down and even start to temporarily rotate in the retrograde direction. However, it is difficult to imagine that such a situation holds much longer than the orbital binary period. Therefore, the possible episodes with negative $\eta$ would somewhat increase the NS equilibrium period $P_{eq}$ which we recalculate at each time step  of our population synthesis calculations. Therefore, we will consider only positive values of $\eta$ and restrict the NS spin rotation by the orbital binary periods, $P^*\le P_b$.



{\underline{\it Orbital eccentricity effect}}.


Here we describe the effect of orbital eccentricity $e$ on the value of the equilibrium NS period at the settling accretion stage. Because of the orbital eccentricity, the spin-up and spin-down torques applied to NS change and should be averaged over the orbital period. In the standard Keplerian problem, the separation between  barycenters of the binary components 
$
r=p/(1+e\cos\theta),
$
where $p=a(1-e^2)$ -- orbital semilatus rectum, $a$ -- 
large semi-major axis of the orbit, $e$ -- orbital eccentricity, $\theta$ --  
 true anomaly. Stellar wind  is assumed to be spherically symmetric and centered at the optical star barycenter. The radius of the optical star is $R_o$, and the stellar wind velocity as a function of distance from the star is normalized to the parabolic velocity at the optical star photosphere, $v_{esc}=\sqrt{2GM_o/R_o}$. It
 can be written as 
$v_w(r)=v_{esc}f(r).$ 
For example, for the radiative-driven accelerating winds from early type stars
$f(r) = \alpha \left(1-R_o/r\right)^\beta$
where $\alpha=0.7...3.8$ and $\beta=0.7...2.0$ are obtained from recent numerical simulations  \citep{2018arXiv180806612V}.

\begin{wrapfigure}{r}{0.5\textwidth}
	\includegraphics[width=0.52\textwidth]{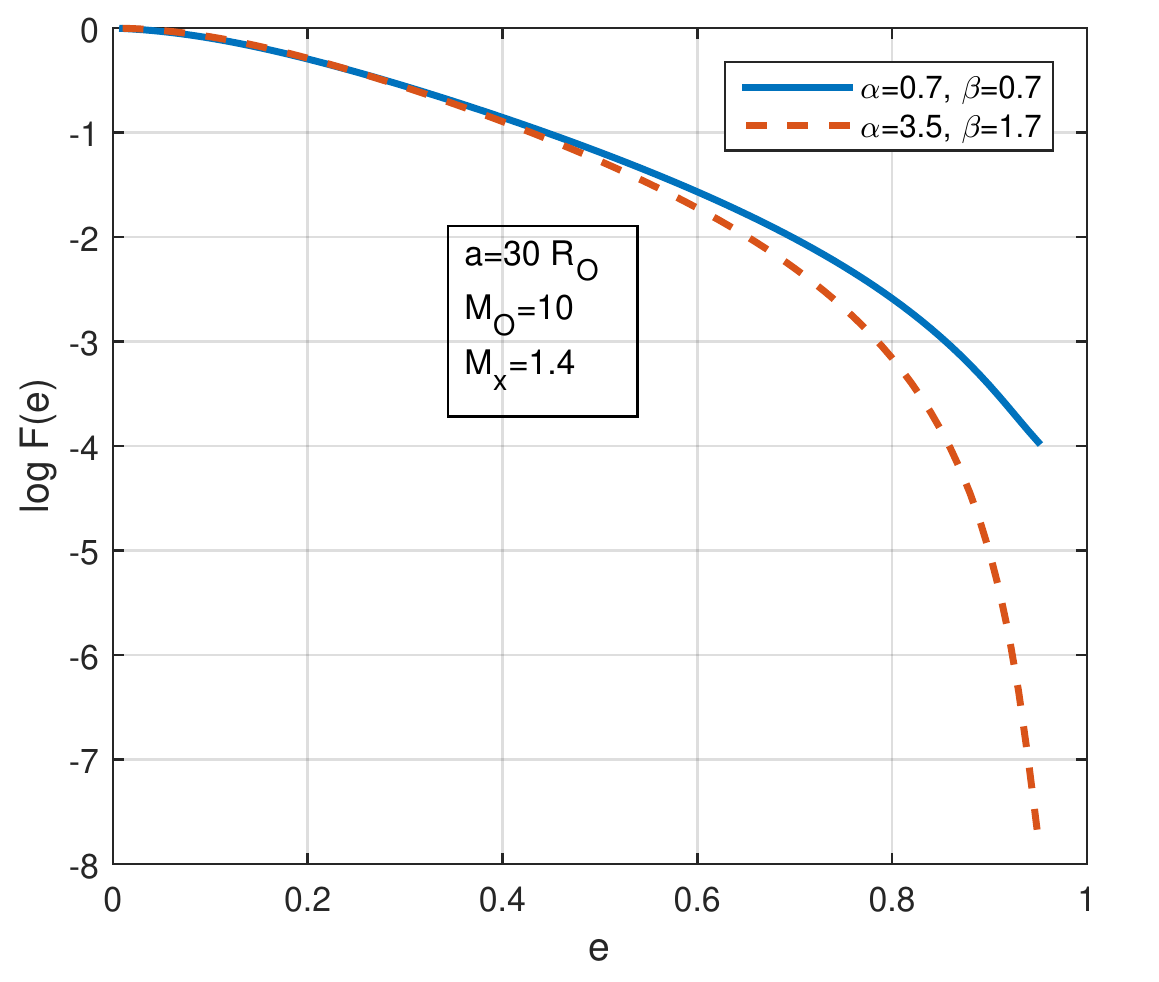}
    \caption{Reduction factor $F(e)$ for the equilibrium NSs periods at the settling accretion stage in the non-circular orbits for the range of radiative stellar wind parameters $\alpha$ and $\beta$.
}
    \label{f:F(e)}
\end{wrapfigure}

In the case of a non-circular orbit, specific angular momentum of captured wind matter is determined by the normal component of the NS orbital velocity
and changes along the orbit: $j(\theta)\!\sim\!v_n(\theta)R_B(\theta)[R_B(\theta)/r(\theta)]$ (here 
$
v_n\!=\!\sqrt{GM/p}(1+e\cos\theta),
$
$M=M_o+M_x$), 
while the gravitational capture Bondi radius $R_B(\theta)$ depends on the module of the sum of the orbital velocity vector 
$\bf{v}_{orb}$ and the wind velocity vector $\bf{v}_w$:
$R_B(\theta)\!\sim\!2GM_x/v(\theta)^2$, where
$
v(\theta)=\sqrt{[v_w(r(\theta))-v_r(\theta)]^2+v_n^2(\theta)}
$
(here the radial component of the orbital velocity is $v_r(\theta)=v_n=\sqrt{GM/p}(1+e\cos\theta)$).

Thus, by averaging the spin-up/spin-down \Eq{e:spinev} 
over orbit $r(\theta)$, we find NS equilibrium period:
\beq{e:spin_eq}
\overline{P_{eq}} = 2\pi \frac{\langle \dot M_B R_A^2\rangle}{\langle \dot M_Bv_nR_B(R_B/r)\rangle}\,.
\eeq
The orbit-averaged value $\overline{P_{eq}} $ 
can be written using the equilibrium period calculated for a circular orbit, $P_{eq}\approx P_b(R_A/R_B)^2$, as 
$\overline{P_{eq}} =F(e)P_{eq}$, where the factor $F(e)<1$ is plotted in Fig. \ref{f:F(e)}. A significant decrease of the equilibrium NSs periods may occur for highly eccentric binaries (e.g., 
Be X-ray binaries, BeXRBs). Indeed, they are located below HMXBs with almost circular orbits in the Corbet diagram. Thus, the eccentricity effect can explain the observed diversity of X-ray pulsar populations in HMXBs produced by different types of supernova noted by \cite{2011Natur.479..372K}.
 
\section{Supercritical accretion onto magnetized NSs}

\begin{figure*}
	\includegraphics[width=1.15\textwidth]{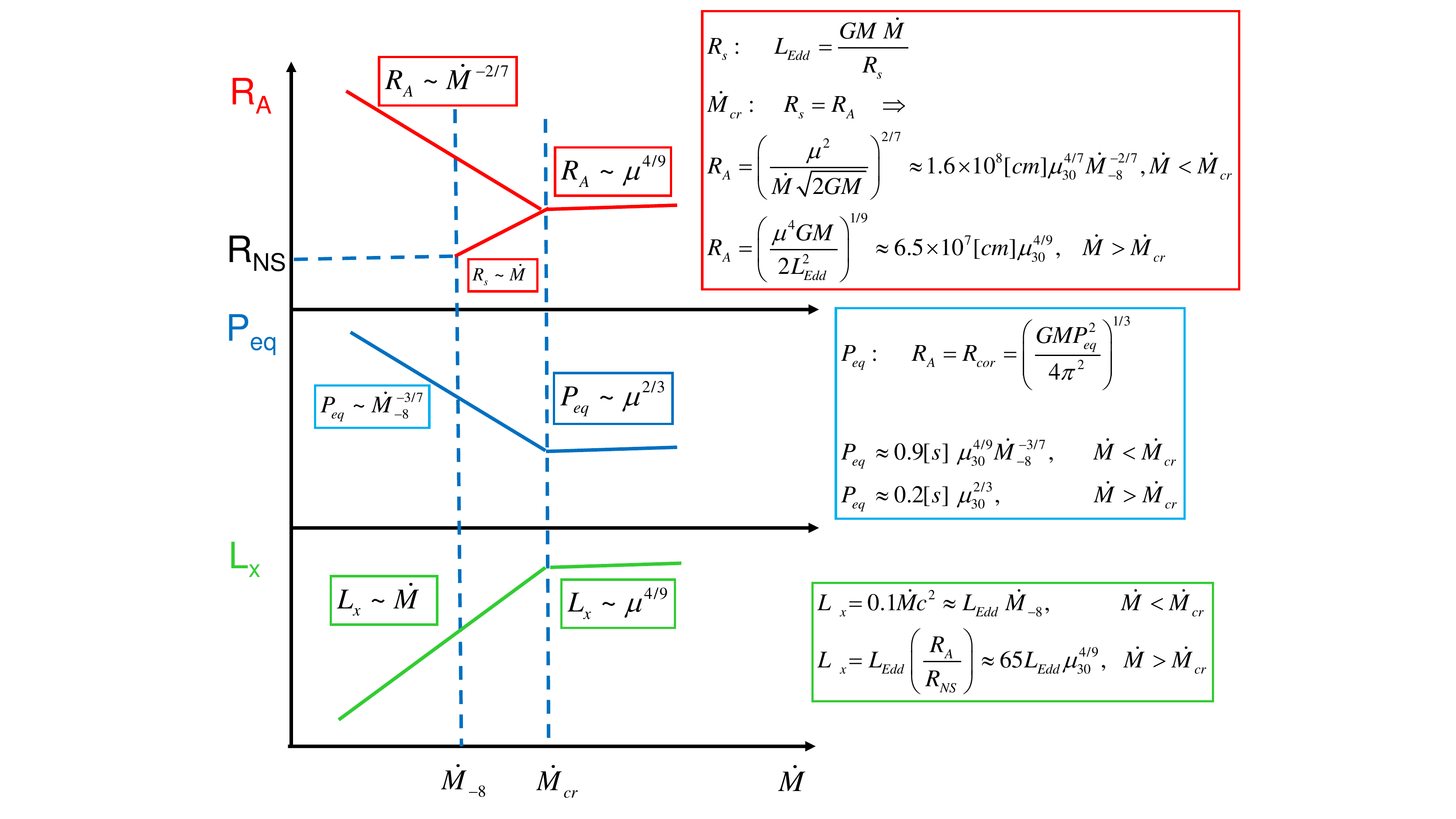}
    \caption{The Alfv\'en radius (top),  equilibrium NS period (middle) and X-ray luminosity (bottom) for supercritical accretion discs around magnetized NSs (see the text).
}
    \label{f:super}
\end{figure*}

In HMXBs with Roche overflow by the optical star, a supercritical disc accretion regime must occur \citep{1973A&A....24..337S}. In this regime, a radiation-driven outflow occurs in the disc starting from the spherization radius 
$R_s=GM_x\dot{M}/L_{Edd}$, where 
$L_{Edd}\simeq 10^{38}(M_x/M_\odot)\mathrm{~[erg\,s^{-1}]}$
is the Eddington luminosity corresponding to 
$\dot{M}_{Edd}\simeq 10^{18}\,\mathrm{[g\,s^{-1}]}$.
 A feature of the supercritical accretion onto a magnetized NS is the appearance of a critical luminosity (mass accretion rate) at which the Alfv\'en radius $R_A$ becomes comparable to the spherization radius $R_s$ as shown in the scheme in Fig. \ref{f:super} (see also \cite{1982SvA....26...54L, 2017MNRAS.468L..59K}), $\dot M_{cr}\simeq 3\times 10^{19}\mu_{30}^{4/9}$~[g s$^{-1}$]. As long as mass accretion rate in the supercritical disc 
increases and  $\dot M_{Edd}<\dot M<\dot M_{cr}$,  NS magnetospheric radius decreases, $R_A\sim \dot M^{-2/7}$, its  equilibrium period decreases, $P^*\sim \dot M ^{-5/7}$, and the X-ray luminosity generated at the NS surface increases, $L_x\sim \dot M$. After reaching $\dot M_{cr}$, NS magnetosphere stops to decrease  because of the wind outflow from $R_s=R_A$ (see Fig. \ref{f:super}), and for $\dot M>\dot M_{cr}$ all three quantities, $R_A$, $P_{eq}$ and $L_x$ become dependent on the NS magnetic field only (the latter to within a logarithmic correction $1+\ln(\dot M/\dot M_{Edd})$, \cite{1973A&A....24..337S}). 

\section{Population synthesis of HMXBs}
\label{s:popsynt}

We use a modified version of the openly available BSE code \citep{2000MNRAS.315..543H,2002MNRAS.329..897H} appended by the block for calculation of  
spin evolution of magnetized neutron stars \citep{2009ARep...53..915L} with account for settling accretion regime \citep[see also][for more detail]{2012MNRAS.424.2265L}. The standard HMXB evolutionary scenario is used \citep{2014LRR....17....3P}. We have used the convolution of the HMXB formation rate after instantaneous star-formation burst with Galactic star-formation rate 
in the Galactic bulge and thin disc 
in the form suggested by \citet{2010A&A...521A..85Y}:
\beq{e:SFR}
\frac{\mathrm{SFR}(t)}{M_\odot\,\mathrm{yr}^{-1}}=\left\{
\begin{array}{lr}
11e^{-\frac{t-t_0}{\tau}}+0.12(t-t_0) & t\ge t_0,\\
0 & t<t_0
\end{array}
\right.
\eeq
with time $t$ in Gyr, $t_0$=4 Gyr, $\tau=9$~Gyr. 
Assumed  Galactic age is 14 Gyr. This model gives total mass of Galactic bulge and thin disk $M_G=7.2\times 10^{10} M_\odot$, which we will use for normalization of our calculation results. 

\begin{figure*}
	\includegraphics[width=0.5\textwidth]{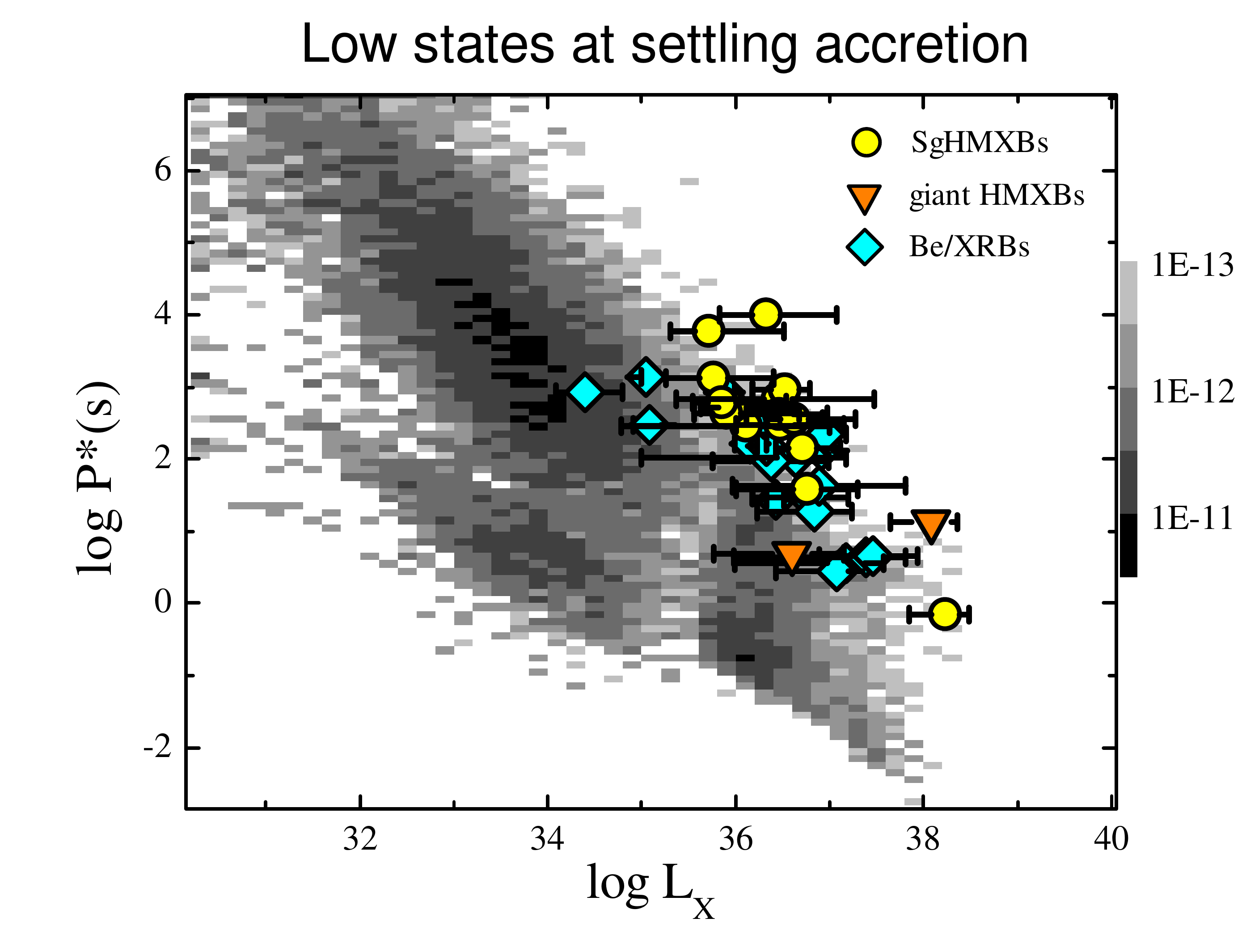}
	\includegraphics[width=0.5\textwidth]{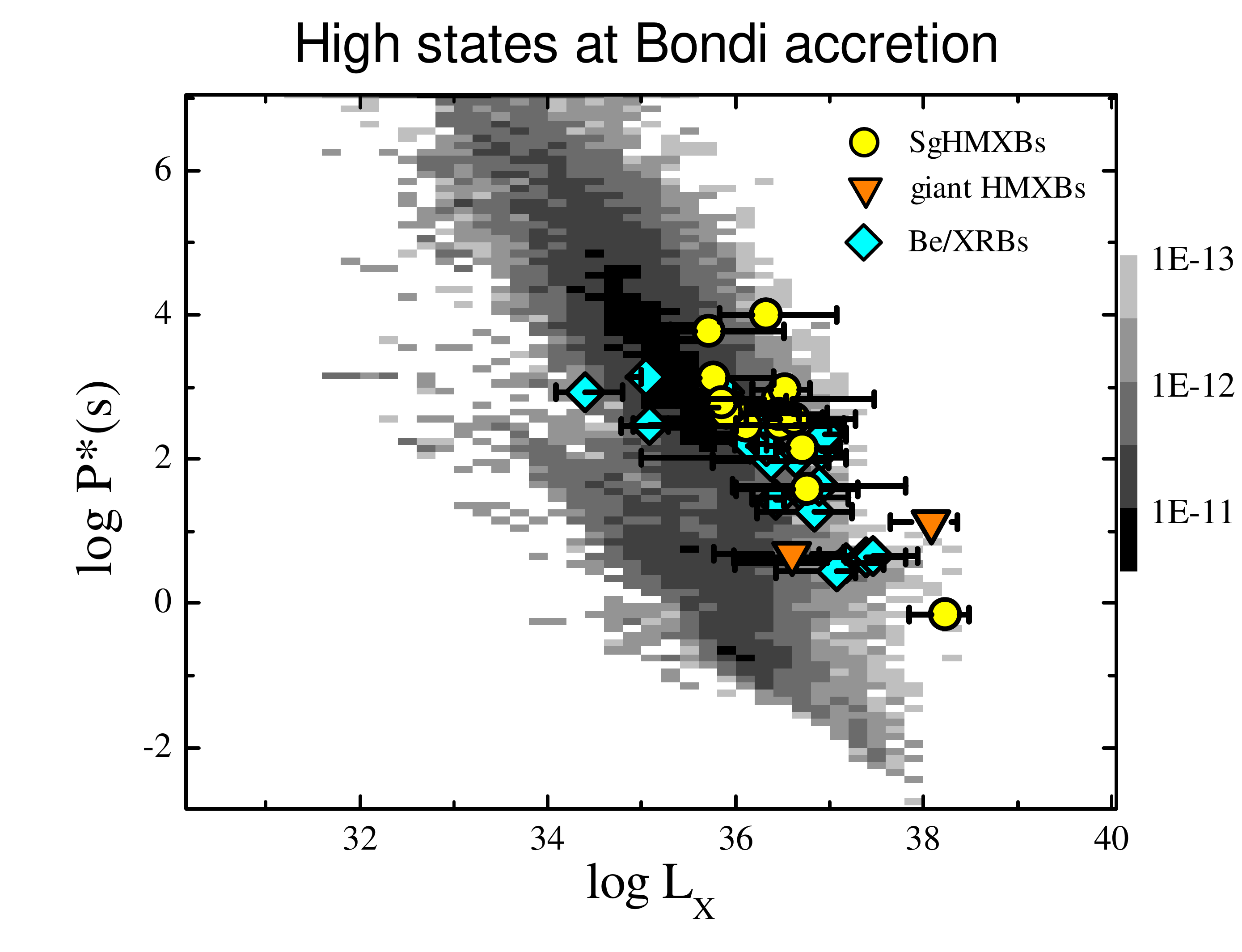}
    \caption{Model $P^*$ -- $L_x$ distribution of HMXBs (in grey scale, normalized to one $M_\odot$ of the Galactic mass) and observed persistent HMXBs (filled circles). Filled diamonds show BeXRBs with  non-circular orbits in outbursts.
}
    \label{f:PLx}
\end{figure*}

To illustrate the effect of different 
regimes 
of wind accretion in HMXBs, in Fig. \ref{f:PLx} 
 we show in grey scale the model Galactic distribution of equilibrium periods $P^*$ of NSs in HMXBs vs. their X-ray luminosity $L_x$ normalized to one $M_\odot$. 
The range of observed $L_x$ of persistent wind-accreting HMXBs is taken from \citet{2018MNRAS.481.2779S}. Outburst X-ray luminosity of BeXRBs with eccentric orbits are shown by filled blue diamonds. 
As discussed above, a wind-accreting NS can reach long spin periods at the subsonic settling accretion stage. However, when the accretion X-ray luminosity exceeds $\sim 4\times 10^{16}$ [erg s$^{-1}$], the supersonic Bondi accretion regime sets in, at which the NS acquires angular momentum from captured wind matter. Therefore, we plot separately the model $P^*-L_x$ distribution for the settling accretion regime (left panel in Fig. \ref{f:PLx}) and the Bondi X-ray luminosity for the same sources ($L_{x,B}=0.1\dot M_B c^2$) (right panel in Fig. \ref{f:PLx}). Clearly, the observed location of persistent HMXBs and BeXRBs on this diagram better corresponds to the Bondi accretion luminosity (but see the discussion of Vela X-1 and GX 301-2 in \citep{2012MNRAS.420..216S}), and the model left and right panels show the expected X-ray luminosity range between settling and Bondi accretion stages. The actual situation may be due to a selection effect: the brightest sources correspond to the more luminous Bondi accretion, and a bulk of possible dimmer sources with about solar X-ray luminosity are yet to be identified in more sensitive observations.

\begin{figure*}
	\includegraphics[width=0.5\textwidth]{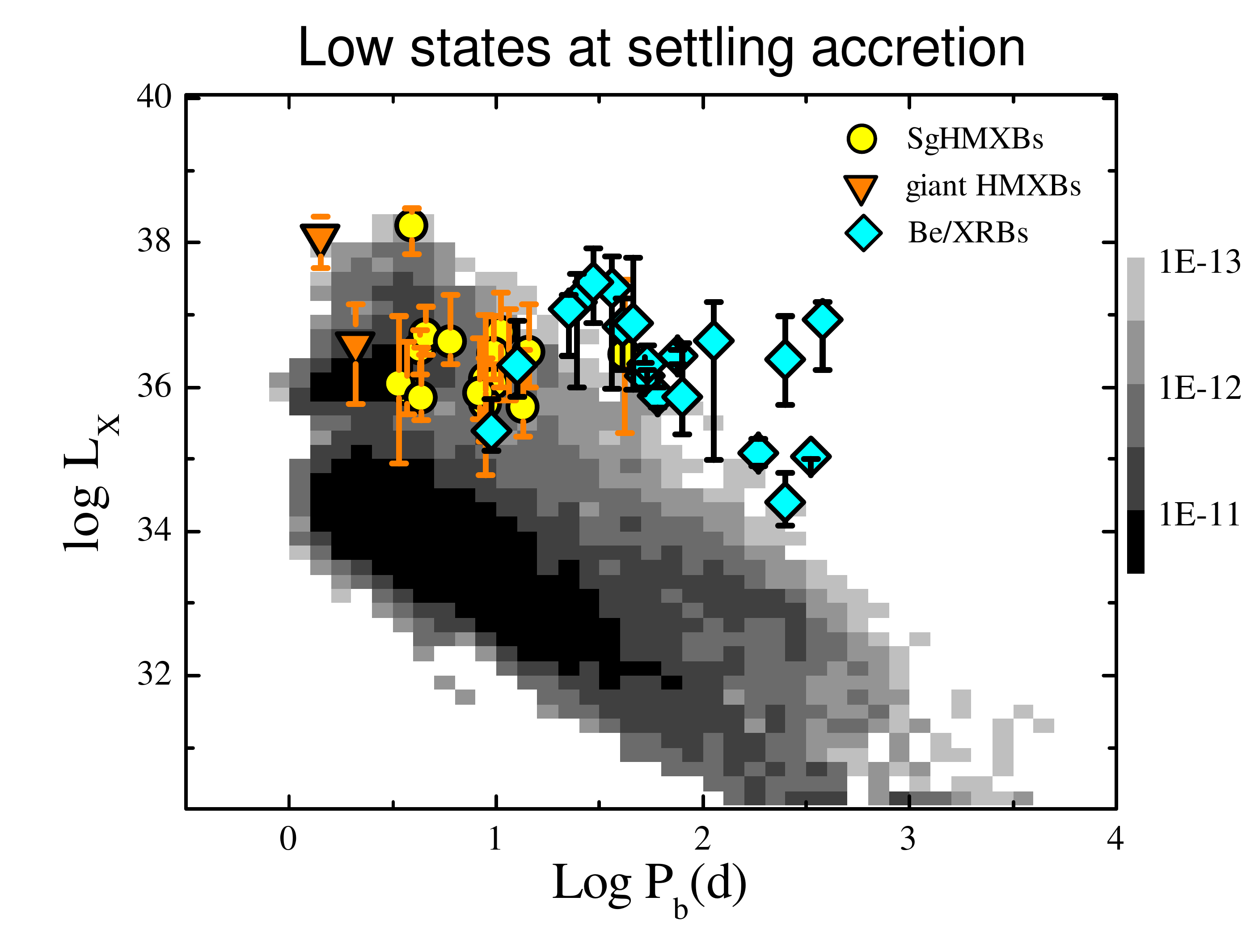}
	\includegraphics[width=0.5\textwidth]{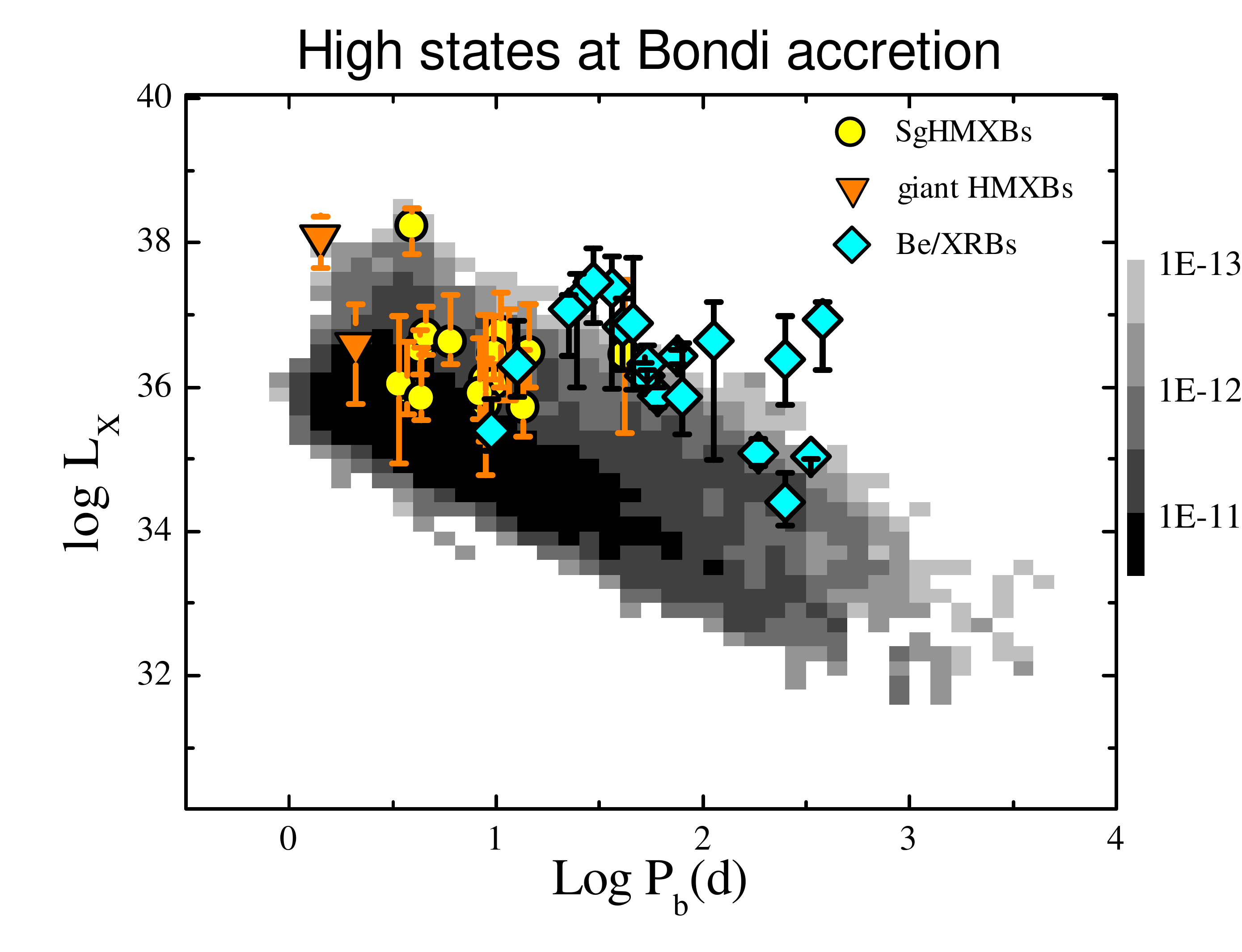}
    \caption{Model X-ray luminosity $L_x$ -- binary orbital period diagram for HMXBs (in grey scale) and observed steady  sources (filled circles). Filled diamonds show BeXRBs with non-circular orbits in outbursts. Data from \citet{2018MNRAS.481.2779S}.
}
    \label{f:LxPorb}
\end{figure*}

\begin{figure*}
	\includegraphics[width=0.5\textwidth]{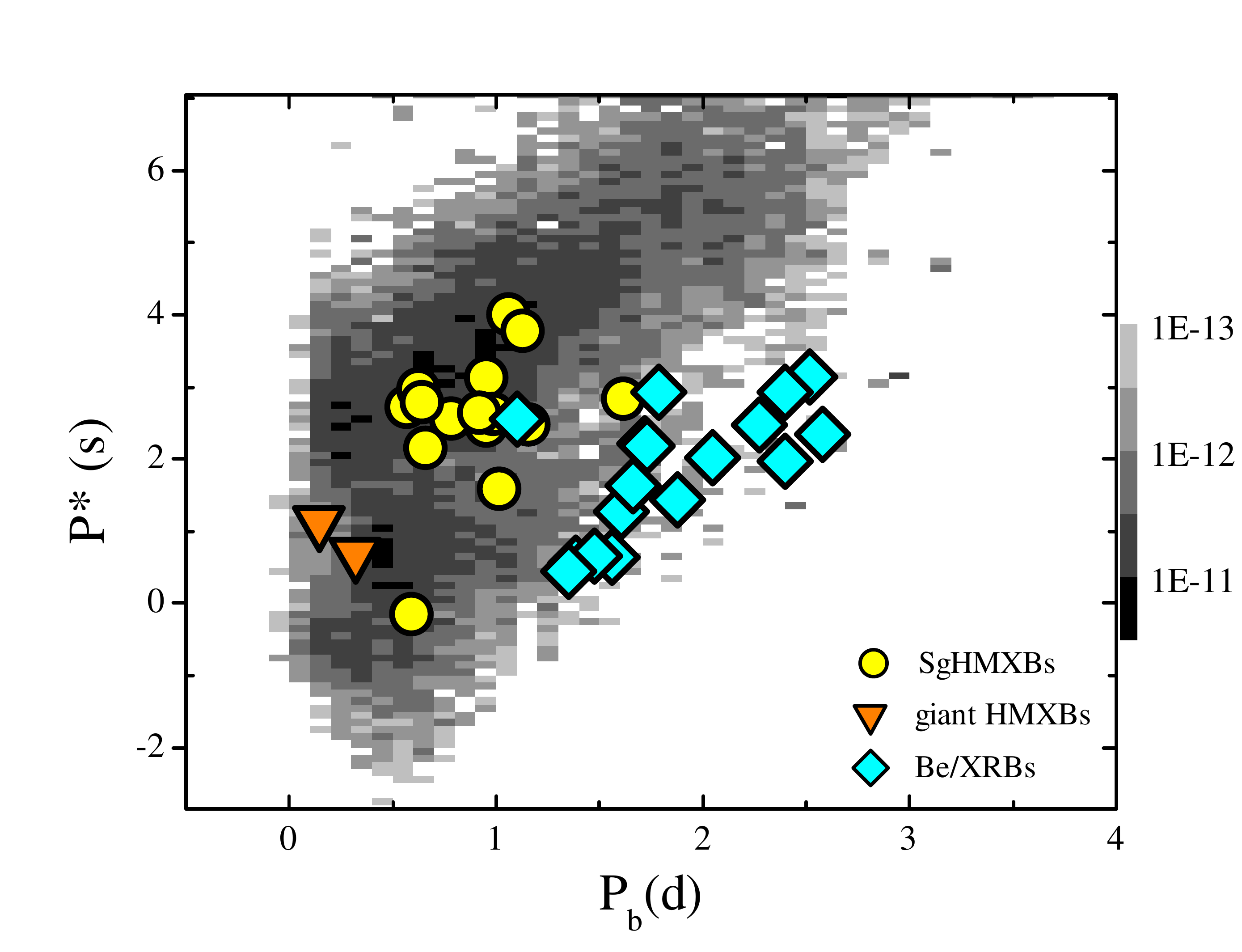}
   \includegraphics[width=0.5\textwidth]{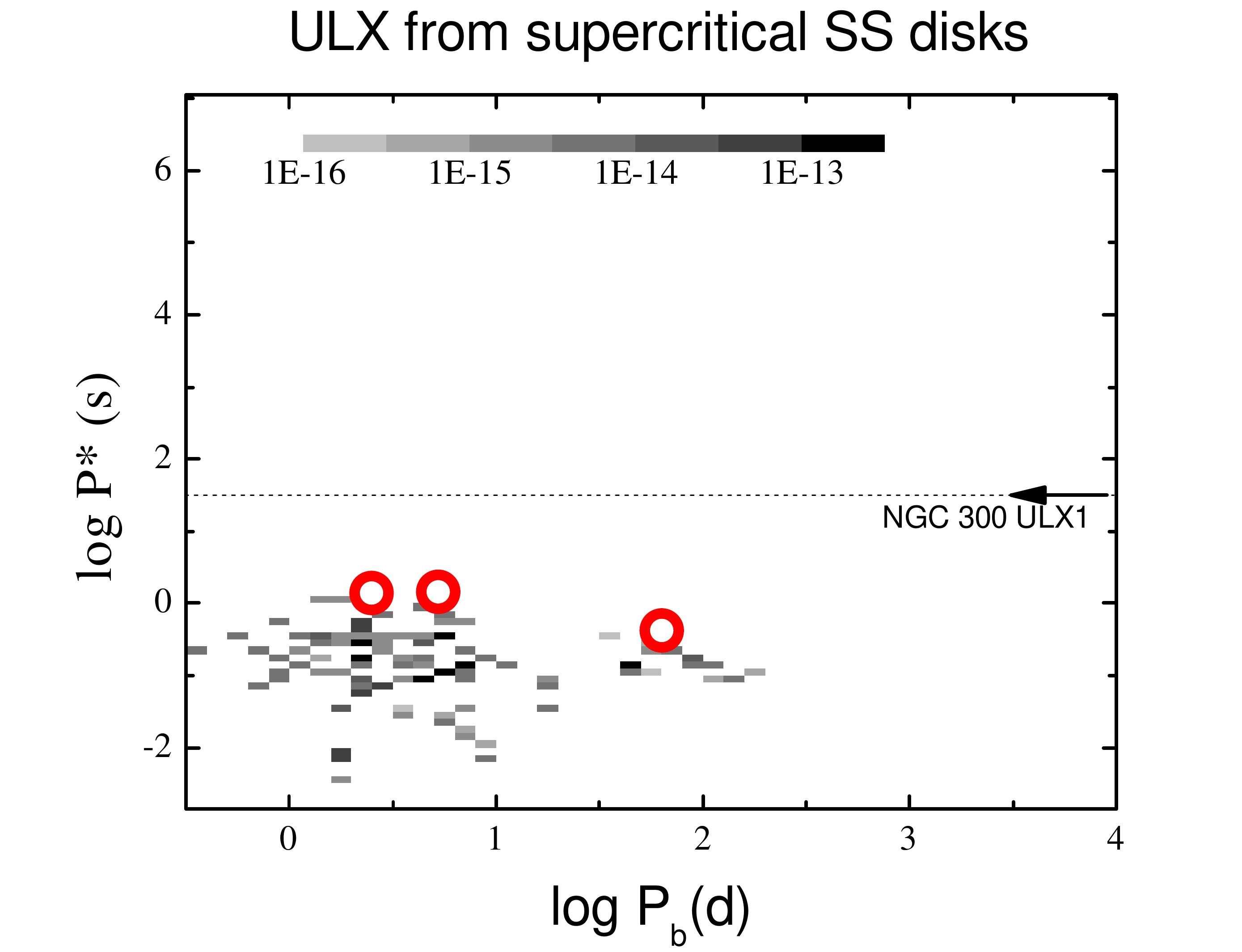}
\caption{Left: model NS spin period $P^*$ -- binary orbital period $P_b$ diagram (the Corbet diagram) for HMXBs (in grey scale) and observed steady  sources (filled circles). Filled diamonds show BeXRBs with non-circular orbits. Data from \citet{2018MNRAS.481.2779S}. Right: the Corbet diagram for NSs at supercritical accretion disc stage. Open red circles show pulsating ULXs,
dashed line indicates $P^*$ of  NGC~300~ULX1.
}
    \label{f:Corbet}
\end{figure*}

\begin{figure*}
	\includegraphics[width=0.5\textwidth]{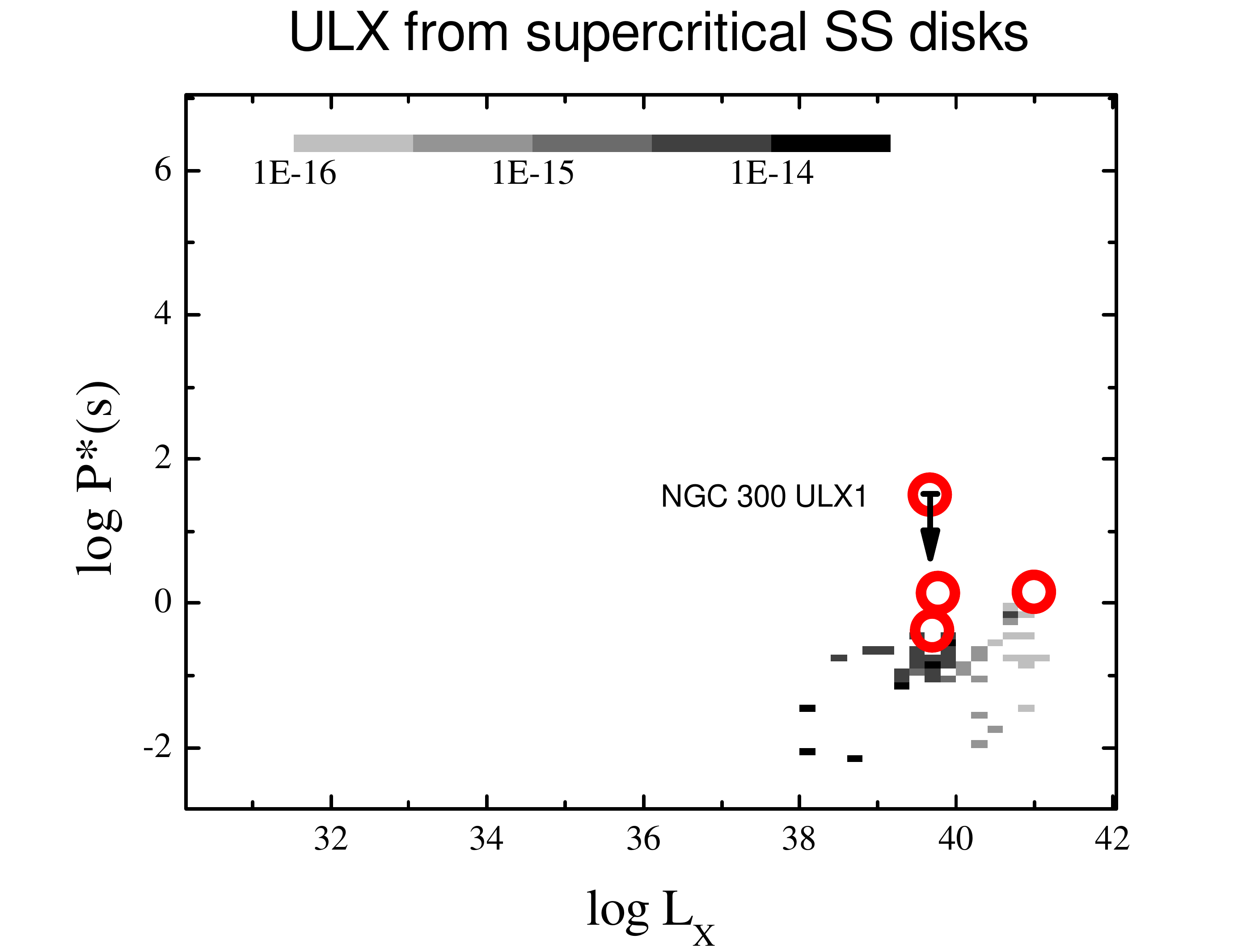}
	\includegraphics[width=0.5\textwidth]{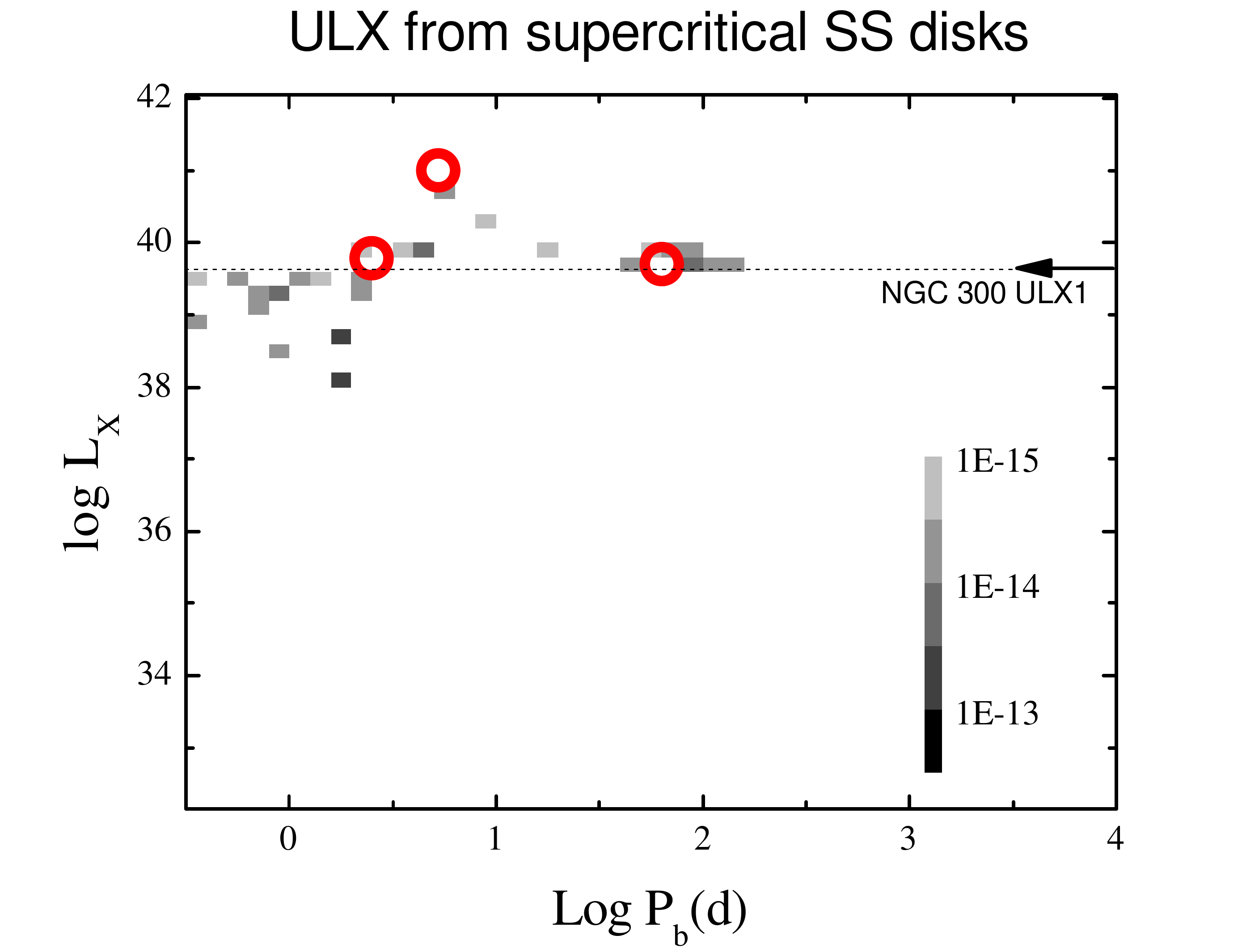}
    \caption{Model $P^*-L_x$ and $L_x-P_b$ diagrams (left and right panels, respectively) for NSs in HMXBs at supercritical accretion disc stage (in grey color, normalized to one $M_\odot$ of Galactic mass). Open red circles show pulsating ULX. NGC 300 ULX X-1 is marked with black arrow.}
     \label{f:ulx}
\end{figure*}

In Fig.~\ref{f:LxPorb} we plot the calculated distribution of sources in the
 X-ray luminosity -- orbital period plane. Like in Fig. \ref{f:PLx}, the left and right panels correspond to settling and Bondi accretion states, respectively. The model $P^*$--$P_b$ distribution (the Corbet diagram) for wind accreting HMXBs is shown in the left panel of Fig. \ref{f:Corbet}. Separately, in the right panel of Fig.~\ref{f:Corbet}, model HMXBs with supercritical Shakura-Sunyaev accretion discs are shown (in grey scale). Observed pulsating ULXs with known orbital periods \citep{2014Natur.514..202B, 2016ApJ...831L..14F, 2017Sci...355..817I, 2017MNRAS.466L..48I} are shown by red open circles.  In 
Fig.~\ref{f:ulx} we plot the model distribution of HMXBs with neutron stars with supercritical accretion discs and observed pulsating ULXs (red open circles). In addition to the observed pulsating ULXs shown in the Corbet diagram (right panel of Fig.~\ref{f:Corbet}), 
we plot there also candidate ULX NGC~300~ULX1 with 
$P^*\simeq$ 31.7\,s, $L_x\simeq 4.7\times 10^{39}$\,[erg s$^{-1}$] \citep{2018MNRAS.476L..45C}.

\section{Discussion and conclusions}
\label{s:concl}

In wind accreting HMXBs with magnetized NSs, two regimes of accretion should be distinguished: the supersonic Bondi-Hoyle-Littleton accretion at high accretion rates, $\dot M\gtrsim 4\times 10^{16}$~[g s$^{-1}$] (high X-ray luminosity $L_x>4\times 10^{36}$~[erg s$^{-1}$]) and subsonic settling accretion (lower $L_x$). 
In the regime of settling accretion, a hot shell is formed around the NS magnetosphere enabling angular momentum transfer to/from the rotating NS, and NS reaches an equilibrium spin period proportional to the orbital binary period $P_b$. Only spin-up of NS is possible 
in the Bondi-Hoyle accretion stage.

{\underline{\it The Corbet diagram for HMXBs}}.
We performed population synthesis calculations of HMXBs with magnetized NSs under standard assumptions on the massive binary star evolution, 
parameters of stellar wind  for early type massive stars and magnetic fields of NSs as inferred from radio pulsars observations. 

Inspection of model distribution $P^*-L_x$ (Fig.~\ref{f:PLx}) suggests that NSs in wind-accreting HMXBs have reached the equilibrium spin period at the early settling accretion stage, but the observed X-ray luminosities better correspond to the Bondi-Hoyle mass accretion rate from the stellar wind (right panel of Fig.~\ref{f:PLx}). The latter holds true for the model $L_x$-$P_b$ distribution of HMXBs (Fig. \ref{f:LxPorb}): the Bondi-Hoyle accretion rates better correspond to observations (right panel of Fig.~\ref{f:LxPorb}), suggesting earlier stage of settling accretion at which NSs have reached the equilibrium spin periods. This conclusion is supported by the model $P^*-P_b$ distribution of HMXBs (the Corbet diagram) shown in Fig.~\ref{f:Corbet}, left panel. 
Parameters of the observed quasi-steady HMXBs (filled yellow circles and triangles) are in good agreement with the model. However, the observed Be X-ray binaries shown by filled blue diamonds 
are located 
 systematically lower in this diagram, most likely, due to the reduction of the equilibrium spin period due to orbital eccentricity by the factor $F(e)$ (see \Eq{e:spin_eq}) presented in Fig.~\ref{f:F(e)}.

{\underline{\it The Corbet diagram for pulsating ULXs}}.
When a massive optical star in HMXB overfills its Roche lobe, a supercritical accretion disc can be formed around a magnetized NS. Then the formulas shown in Fig.~\ref{f:super}  should be used to calculate the NS equilibrium spin period $P^*$ and X-ray luminosity $L_x$ (in the case 
where $\dot M$ in the disc exceeds $\dot M_{cr}$ defined in Sec.~3). Population synthesis calculations of the HMXBs that during later evolution produced objects hosting supercritical accretion discs shown in Fig.~\ref{f:Corbet}, right panel (the Corbet diagram for ULXs) and Fig.~\ref{f:ulx} 
suggest that pulsating ULXs with known spin $P^*$ and orbital periods $P_b$ fall almost precisely into the expected distribution (shown in grey scale normalized to one $M_\odot$ of the Galactic mass). We stress that in these calculations, we have not done any special assumptions about NS properties (e.g., the initial magnetic field distribution, which is assumed to be a Gaussian centered at $10^{12}$~G) or binary stellar evolution, i.e. the wind-accreting HMXBs and pulsating ULXs represent single
 evolutionary related population. Here we stress that the equilibrium NS spin periods and luminosities at supercritical accretion stages depend only on the NS magnetic field (see formulas in Fig. \ref{f:super}). Therefore, the observed dispersion in spin periods of pulsating ULXs around $\sim 1$ sec reflects the dispersion of magnetic fields of superaccreting NSs in HMXBs. It is clear that to explain these spin periods, no hypothesis of unusual magnetic fields of NSs is needed. 

The 31.7-s NGC 300 ULX1 shown as the upper circle in the left panel of
Fig.~\ref{f:ulx}, 
 is worth special mentioning. Its record-high observed spin-up rate \citep{2018MNRAS.476L..45C} $\dot P\sim 5\times 10^{-7}$[s~s$^{-1}$] suggests a NS spinning-up towards equilibrium. Our model presented in Fig. \ref{f:super} immediately implies that for $\dot M> \dot M_{cr}$ (which is the case for the observed X-ray luminosity $\sim 5 \times 10^{39}$ [erg s$^{-1}$]), the spin-up rate far from equilibrium is $\dot P=P^2/(2\pi I)\sqrt{GM_x R_A}\approx P^2/(2\pi I)\sqrt{GM_x\times 6.5\cdot 10^7(L_x/(65 L_{Edd}))}\simeq 5\times 10^{-7}$ [s s$^{-1}$], in perfect agreement with observations.

\underline{\it{Acknowledgements}}. KP acknowledges grant RSF 16-12-10519 for partial support (Section 1-3). The work of AK (Section 4) is supported by RSF grant No. 14-12-00146.  

\bibliographystyle{iau2}
\bibliography{hmxb}

\begin{thebibliography}{30}
\providecommand{\natexlab}[1]{#1}

\bibitem[{{Arons} \& {Lea}(1976)}]{1976ApJ...207..914A}
{Arons}, J. \& {Lea}, S.M. 1976, \emph{\apj}, 207, 914

\bibitem[{{Arons} \& {Lea}(1980)}]{1980ApJ...235.1016A}
{Arons}, J. \& {Lea}, S.M. 1980, \emph{\apj}, 235, 1016

\bibitem[{{Bachetti} et~al.(2014)}]{2014Natur.514..202B}
{Bachetti}, M. et~al. 2014, \emph{\nat}, 514, 202

\bibitem[{{Burnard} et~al.(1983){Burnard}, {Arons}, \&
  {Lea}}]{1983ApJ...266..175B}
{Burnard}, D.J., {Arons}, J., \& {Lea}, S.M. 1983, \emph{\apj}, 266, 175

\bibitem[{{Carpano} et~al.(2018){Carpano}, {Haberl}, {Maitra}, \&
  {Vasilopoulos}}]{2018MNRAS.476L..45C}
{Carpano}, S., {Haberl}, F., {Maitra}, C., \& {Vasilopoulos}, G. 2018,
  \emph{\mnras}, 476, L45

\bibitem[{{Chakrabarty} et~al.(1997)}]{1997ApJ...481L.101C}
{Chakrabarty}, D. et~al. 1997, \emph{\apjl}, 481, L101

\bibitem[{{de Val-Borro} et~al.(2017){de Val-Borro}, {Karovska}, {Sasselov}, \&
  {Stone}}]{2017MNRAS.468.3408D}
{de Val-Borro}, M., {Karovska}, M., {Sasselov}, D.D., \& {Stone}, J.M. 2017,
  \emph{\mnras}, 468, 3408

\bibitem[{{F{\"u}rst} et~al.(2016)}]{2016ApJ...831L..14F}
{F{\"u}rst}, F. et~al. 2016, \emph{\apjl}, 831, L14

\bibitem[{{Gonz{\'a}lez-Gal{\'a}n} et~al.(2012)}]{2012A&A...537A..66G}
{Gonz{\'a}lez-Gal{\'a}n}, A. et~al. 2012, \emph{\aap}, 537, A66

\bibitem[{{Ho} et~al.(1989){Ho}, {Taam}, {Fryxell}, {Matsuda}, \&
  {Koide}}]{1989MNRAS.238.1447H}
{Ho}, C., {Taam}, R.E., {Fryxell}, B.A., {Matsuda}, T., \& {Koide}, H. 1989,
  \emph{\mnras}, 238, 1447

\bibitem[{{Hunt}(1971)}]{1971MNRAS.154..141H}
{Hunt}, R. 1971, \emph{\mnras}, 154, 141

\bibitem[{{Hurley} et~al.(2000){Hurley}, {Pols}, \&
  {Tout}}]{2000MNRAS.315..543H}
{Hurley}, J.R., {Pols}, O.R., \& {Tout}, C.A. 2000, \emph{\mnras}, 315, 543

\bibitem[{{Hurley} et~al.(2002){Hurley}, {Tout}, \&
  {Pols}}]{2002MNRAS.329..897H}
{Hurley}, J.R., {Tout}, C.A., \& {Pols}, O.R. 2002, \emph{\mnras}, 329, 897

\bibitem[{{Illarionov} \& {Sunyaev}(1975)}]{1975A&A....39..185I}
{Illarionov}, A.F. \& {Sunyaev}, R.A. 1975, \emph{\aap}, 39, 185

\bibitem[{{Israel} et~al.(2017{\natexlab{a}})}]{2017Sci...355..817I}
{Israel}, G.L. et~al. 2017{\natexlab{a}}, \emph{Science}, 355, 817

\bibitem[{{Israel} et~al.(2017{\natexlab{b}})}]{2017MNRAS.466L..48I}
{Israel}, G.L. et~al. 2017{\natexlab{b}}, \emph{\mnras}, 466, L48

\bibitem[{{King} et~al.(2017){King}, {Lasota}, \&
  {Klu{\'z}niak}}]{2017MNRAS.468L..59K}
{King}, A., {Lasota}, J.P., \& {Klu{\'z}niak}, W. 2017, \emph{\mnras}, 468, L59

\bibitem[{{Knigge} et~al.(2011){Knigge}, {Coe}, \&
  {Podsiadlowski}}]{2011Natur.479..372K}
{Knigge}, C., {Coe}, M.J., \& {Podsiadlowski}, P. 2011, \emph{\nat}, 479, 372

\bibitem[{{Lipunov}(1982)}]{1982SvA....26...54L}
{Lipunov}, V.M. 1982, \emph{SvA Lett.}, 26, 54

\bibitem[{{Lipunov} et~al.(2009){Lipunov}, {Postnov}, {Prokhorov}, \&
  {Bogomazov}}]{2009ARep...53..915L}
{Lipunov}, V.M., {Postnov}, K.A., {Prokhorov}, M.E., \& {Bogomazov}, A.I. 2009,
  \emph{Astronomy Reports}, 53, 915

\bibitem[{{Liu} et~al.(2017){Liu}, {Stancliffe}, {Abate}, \&
  {Matrozis}}]{2017ApJ...846..117L}
{Liu}, Z.W., {Stancliffe}, R.J., {Abate}, C., \& {Matrozis}, E. 2017,
  \emph{\apj}, 846, 117

\bibitem[{{L{\"u}} et~al.(2012)}]{2012MNRAS.424.2265L}
{L{\"u}}, G.L. et~al. 2012, \emph{\mnras}, 424, 2265

\bibitem[{{Postnov} \& {Yungelson}(2014)}]{2014LRR....17....3P}
{Postnov}, K.A. \& {Yungelson}, L.R. 2014, \emph{Living Reviews in Relativity},
  17, 3

\bibitem[{{Shakura} \& {Postnov}(2017)}]{2017arXiv170203393S}
{Shakura}, N. \& {Postnov}, K. 2017, \emph{ArXiv e-prints}

\bibitem[{{Shakura} et~al.(2012){Shakura}, {Postnov}, {Kochetkova}, \&
  {Hjalmarsdotter}}]{2012MNRAS.420..216S}
{Shakura}, N., {Postnov}, K., {Kochetkova}, A., \& {Hjalmarsdotter}, L. 2012,
  \emph{\mnras}, 420, 216

\bibitem[{{Shakura} et~al.(2018){Shakura}, {Postnov}, {Kochetkova}, \&
  {Hjalmarsdotter}}]{APA}
{Shakura}, N., {Postnov}, K., {Kochetkova}, A., \& {Hjalmarsdotter}, L. 2018,
  in: N.~Shakura, ed., \emph{Accretion Processes in Astrophysics}, chap.~8,
  Springer:

\bibitem[{{Shakura} \& {Sunyaev}(1973)}]{1973A&A....24..337S}
{Shakura}, N.I. \& {Sunyaev}, R.A. 1973, \emph{\aap}, 24, 337

\bibitem[{{Sidoli} \& {Paizis}(2018)}]{2018MNRAS.481.2779S}
{Sidoli}, L. \& {Paizis}, A. 2018, \emph{\mnras}, 481, 2779

\bibitem[{{Vink}(2018)}]{2018arXiv180806612V}
{Vink}, J.S. 2018, \emph{ArXiv e-prints}, 1808.06612

\bibitem[{{Yu} \& {Jeffery}(2010)}]{2010A&A...521A..85Y}
{Yu}, S. \& {Jeffery}, C.S. 2010, \emph{\aap}, 521, A85

\end{thebibliography}
\end{document}